\documentclass[aps,prc,twocolumn,showpacs,superscriptaddress,floatfix,10pt]{revtex4-1}
\usepackage{graphicx}
\usepackage{dcolumn}
\usepackage{bm}
\usepackage{amsfonts}
\usepackage{amssymb}
\usepackage{amsmath}

\begin{document}

\title{Quantal description of nucleon exchange in stochastic mean-field approach}
\author{S. Ayik}
\affiliation{Physics Department, Tennessee Technological University, Cookeville, Tennessee 38505, USA}
\author{O. Yilmaz}
\affiliation{Physics Department, Middle East Technical University, 06531, Ankara, Turkey}
\author{B. Yilmaz}
\affiliation{Physics Department, Faculty of Sciences, Ankara University, 06100, Ankara, Turkey}
\author{A. S. Umar}
\affiliation{Department of Physics and Astronomy, Vanderbilt University, Nashville, Tennessee 37235, USA}
\author{A. Gokalp}
\affiliation{Department of Physics, Bilkent University, 06800, Ankara, Turkey}
\author{G. Turan}
\affiliation{Physics Department, Middle East Technical University, 06531, Ankara, Turkey}
\author{D. Lacroix}
\affiliation{Institut de Physique Nucl\'eaire, IN2P3-CNRS, Universit\'e Paris-Sud, F-91406 Orsay Cedex, France}

\date{\today}

\begin{abstract} 
Nucleon exchange mechanism is investigated in central collisions of symmetric heavy-ions in the basis of the stochastic mean-field approach. Quantal diffusion coefficients for nucleon exchange are calculated by including non-Markovian effects and shell structure. Variances of fragment mass distributions are calculated in central collisions of ${}^{40}$Ca + ${}^{40}$Ca, ${}^{48}$Ca + ${}^{48}$Ca and ${}^{56}$Ni + ${}^{56}$Ni systems.
\end{abstract}


\maketitle

\section{Introduction}
The standard mean-field theory provides a good approximation for the average evolution of the nuclear collective motion at low energies, but severely underestimates the fluctuation of collective variables~\cite{Si12,Ne82}. Considerable effort has been made to extend the time-dependent Hartree-Fock (TDHF) theory beyond the mean-field approximation~\cite{Bal84,Ayik88,Rand90,Abe96,Lac04,Sim10,TU02}. The Stochastic Mean-Field (SMF) approach goes beyond the standard mean-field description by incorporating the quantal and thermal fluctuations in the initial state~\cite{Ayik08}. The initial state fluctuations, which can be specified in a suitable manner, are incorporated into the dynamics by generating an ensemble of single-particle density matrix according to the fluctuations in the initial state. In a number of applications, it was illustrated that the SMF approach provides very good approximation for exact quantal evolution of the many-body systems at low energies, where collisional dissipation mechanism does not play an important role~\cite{Lac12,Lac13,Yil14}. For a description of the approach and its various applications we refer to~\cite{Lac14,Yilmaz14}.

Recently, we investigated nucleon exchange mechanism in the central~\cite{Ayik09,Washi09,Yil11} and off-central heavy-ion collisions~\cite{Rand78} by employing the SMF approach in the semi-classical approximation and ignoring memory effect in the diffusion process. Transport coefficients extracted from the SMF approach in the semi-classical limit have similar form as in the empirical nucleon exchange model~\cite{Yilmaz2014}, but provide a more refined description of nucleon exchange mechanism. In the present work, we study nucleon exchange mechanism in fully quantal framework of the SMF approach, also incorporating the memory effect in the diffusion process, and compare the results with the semi-classical approximation. In this investigation, for simplicity, we consider the central collisions of symmetric heavy-ions at energies below the fusion barrier. In section 2, we present formal description of the nucleon diffusion in the quantal framework of the SMF approach. In section 3, we carry out calculations of variances of fragment mass distributions in several symmetric collisions. Conclusions are given in section 4.

\section{Quantal diffusion}
The standard TDHF provides a deterministic description of a collision process, i.e. the system evolves from a specified initial condition to a single final state~\cite{Si12}. On the other hand, in the SMF approach the initial condition is specified by a distribution function characterizing the quantal and thermal fluctuations of the initial state. The initial fluctuations are incorporated into the dynamics by generating an ensemble of the single particle density matrices. The expectation values of the observables are evaluated by carrying out averages over the generated ensemble. In a single event labeled by $\lambda $, the single-particle density matrix is determined by evolving the single-particle wave functions $\Phi _{j}^{\lambda } (\vec{r},t)$ according to the self-consistent Hamiltonian in that event. Consequently, in a given event, nucleon density and current density are given by
\begin{eqnarray} \label{eq1} 
\rho ^{\lambda } (\vec{r},t)=\sum _{ij}\Phi _{j}^{*\lambda } (\vec{r},t)\rho _{ji}^{\lambda }  \Phi _{i}^{\lambda } (\vec{r},t), 
\end{eqnarray} 
and
\begin{eqnarray} \label{eq2} 
\vec{j}^{\lambda } (\vec{r},t)=\sum _{ij}\frac{\hbar }{2im}&&\left[\Phi _{j}^{*\lambda } (\vec{r},t)\vec{\nabla }\Phi _{i}^{\lambda } (\vec{r},t)\right.\nonumber\\
&&\left.-\Phi _{i}^{\lambda } (\vec{r},t)\vec{\nabla }\Phi _{j}^{*\lambda } (\vec{r},t)\right] \rho _{ji}^{\lambda },
\end{eqnarray} 
where labels $(i,j)$ indicate a complete set of quantum numbers for specifying single-particle wave functions. In these expressions, elements of density matrix $\rho _{ji}^{\lambda } $ are uncorrelated random Gaussian numbers with zero mean values $\overline{\rho _{ji}^{\lambda } }=0$ and variances determined by
\begin{eqnarray} \label{eq3} 
\overline{\delta \rho _{ji}^{\lambda } \delta \rho _{i'j'}^{\lambda } }=\frac{1}{2} \delta _{ii'} \delta _{jj'} \left[n_{i} (1-n_{j} )+n_{j} (1-n_{i} )\right]. 
\end{eqnarray} 
The average occupation numbers $n_{j} $ are zero or one at zero temperature, and specified by the Fermi-Dirac distribution at finite temperatures~\cite{Ayik08}. In this expression and below, the bar over the quantities indicates the average values over the ensemble generated in the simulation. The current density for each event obeys the continuity equation,  
\begin{eqnarray} \label{eq4} 
\frac{\partial }{\partial t} \rho ^{\lambda } (\vec{r},t)+\vec{\nabla }\cdot \vec{j}^{\lambda } (\vec{r},t)=0. 
\end{eqnarray} 
In deep-inelastic collisions, since binary character of the system is maintained, a set of macroscopic variables can be defined with the help of the window between the colliding ions. In the central collisions of symmetric systems, the collision geometry is rather simple, and the window is located at the origin of the center of mass frame and it is perpendicular to the collision direction. In this work, we do not differentiate between protons and neutrons, we consider only total nucleon  diffusion. We can define the mass number of the projectile-like fragments (or target-like fragments) in each event by integrating over the nucleon density on the right side (or left-side) of the window as,
\begin{eqnarray} \label{eq5} 
A_{p}^{\lambda } (t)=\int d^{3} r\theta (x-x_{0} ) \rho ^{\lambda } (\vec{r},t), 
\end{eqnarray} 
where $x_{0}=0$ denotes the location of the window, which is taken to be at the origin. According to the SMF approach, the mass number of the projectile-like fragment (or target-like) follows a stochastic evolution according to the Langevin equation~\cite{Gard91,Weiss}, 
\begin{eqnarray} \label{eq6} 
\frac{d}{dt} A_{p}^{\lambda } (t)&=&\int dydz j_{x}^{\lambda } (\vec{r},t)|_{x=x_{0} }\nonumber\\
 &=&v_{A} (A_{p}^{\lambda } ,t)+\delta v_{A}^{\lambda } (t), 
\end{eqnarray} 
where $j_{x}^{\lambda } (\vec{r},t)$ denotes component of the current density along the collision direction, which is taken to be as the x-component. The fluctuations of the nucleon flux across the window in general has two contributions. One contribution arises from the event dependence of the nucleon drift coefficient $v_{A} (A_{p}^{\lambda } ,t)$ through the fluctuating mass number. The other part of the fluctuations arises from the elements $\rho _{ji}^{\lambda } $ of the initial density matrix. In this analysis, we consider small amplitude fluctuations and ignore the event dependence of the drift coefficient. Therefore, in Eq.~\eqref{eq6} we replace the fluctuating nucleon drift coefficient by its mean value, $v_{A} (A_{p}^{\lambda } ,t)\approx v_{A} (A_{p} ,t)\equiv v_{A} (t)$,
\begin{eqnarray} \label{eq7} 
v_{A} (t)=\frac{\hbar }{2im} \int dydz \sum _{j}&&\left[\Phi _{j}^{*} (\vec{r},t)\vec{\nabla }\Phi _{j} (\vec{r},t)\right.\nonumber\\
&&\left. -\Phi _{j} (\vec{r},t)\vec{\nabla }\Phi _{j}^{*} (\vec{r},t)\right] _{x=0} n_{j}.      
\end{eqnarray} 
The mean value of the drift is determined by the net nucleon flux across the window between colliding ions. Since in the collisions of symmetric systems, net flux across the window is zero, the mean value of the nucleon drift vanishes, $v_{A} (t)=0$. The fluctuating part of the nucleon flux which arises from the initial fluctuations is given in terms of the elements $\rho _{ji}^{\lambda }$ of the initial density matrix as,
\begin{eqnarray} \label{eq8} 
\delta v_{A}^{\lambda } (t)=\frac{\hbar }{2im} \int dydz \sum _{ij}&&\left[\Phi _{j}^{*} (\vec{r},t)\vec{\nabla }\Phi _{i} (\vec{r},t)\right.\nonumber\\
&&\!\!\!\!-\left.\Phi _{i} (\vec{r},t)\vec{\nabla }\Phi _{j}^{*} (\vec{r},t)\right]_{x=0}\!\!  \delta \rho _{ji}^{\lambda }.  
\end{eqnarray} 
According to the Langevin description, the fluctuating flux acts as a stochastic force on the mass number. Using Eq.~\eqref{eq3} at zero temperature, it is possible to express the correlation function of the fluctuating nucleon flux as,
\begin{eqnarray} \label{eq9} 
\overline{\delta v_{A}^{\lambda } (t)\delta v_{A}^{\lambda } (t')}=\sum _{p}G_{p} (t,t') +\int G_{p} (t,t' )\rho (\varepsilon _{p} )d\varepsilon _{p}.  
\end{eqnarray} 
Here, the summation $p$ in the first term is over the discrete negative energy particle sates, while the integral in the second term is carried out over the positive energy continuum states. The density of states of the continuum states is indicated by $\rho (\varepsilon _{p} )$ and the quantity $G_{p} (t,t')$ is given by
\begin{eqnarray} \label{eq10} 
G_{p} (t,t')=\left(\frac{\hbar }{2m} \right)^{2} \frac{1}{2} \sum _{h}&&\left[A_{ph} (t)\cdot A_{ph}^{*} (t')\frac{}{}\right.\nonumber\\
&&\left.\frac{}{}+A_{ph}^{*} (t)\cdot A_{ph} (t')\right].  
\end{eqnarray} 
In this expression, the summation $h$ runs over occupied hole states, and the particle-hole elements of the matrix $A(t)$ are given by, 
\begin{eqnarray} \label{eq11} 
A_{ph} (t)=\int dydz &&\left[\Phi _{p}^{*} (\vec{r},t)\nabla _{x} \Phi _{h} (\vec{r},t)\right.\nonumber\\
&&\left.-\Phi _{h} (\vec{r},t)\nabla _{x} \Phi _{p}^{*} (\vec{r},t)\right]_{x=0}.            
\end{eqnarray} 
The variance of the mass distribution is defined as $\sigma _{AA}^{2} (t)=\overline{\delta A_{P}^{\lambda } (t)\delta A_{P}^{\lambda } (t)}$. For small amplitude fluctuations neglecting the effect arising from the fluctuations in nucleon drift coefficient in Eq.~\eqref{eq6}, we can deduce the following equation for the variance of the fragment distribution, 
\begin{eqnarray} \label{eq12} 
\frac{d}{dt} \sigma _{AA}^{2} (t)=2D_{AA} (t). 
\end{eqnarray} 
Here, the quantal and memory dependent diffusion coefficient for nucleon exchange is determined by the correlation function of the stochastic part of the nucleon flux according to,
\begin{eqnarray} \label{eq13} 
D_{AA} (t)=\int _{0}^{t}dt' \overline{\delta v_{A}^{\lambda } (t)\delta v_{A}^{\lambda } (t')}. 
\end{eqnarray} 
As can be seen from Eq.(9), the nucleon diffusion coefficient is given as the sum of proton and neutron diffusion coefficients, $D_{AA}=D_{ZZ}+D_{NN}$, and 
there is no mixed diffusion coefficient $D_{ZN}$ as a result of the independent nature of the nucleon exchange.  

\section{Results}
In the previous semi-classical calculations~\cite{Ayik09,Washi09,Yil11,Rand78}, we employed the TDHF code of Kim~\textit{et al.}~\cite{Kim}. In this work, we carry out calculations of the quantal diffusion coefficients for nucleon exchange in the central collisions of ${}^{40}$Ca + ${}^{40}$Ca, ${}^{48}$Ca + ${}^{48}$Ca and ${}^{56}$Ni + ${}^{56}$Ni by employing the TDHF code of Umar \textit{et al.}~\cite{Umar05,US91}, and compare the quantal diffusion coefficients with their semi-classical values. 
Original version of this code calculates only the time dependent occupied wave functions. In order to determine the quantal diffusion coefficient, we extended the code to calculate the time-dependent unoccupied single-particle wave functions in addition to the occupied hole states.
In practice, 3000$-$4000 positive energy states have been used in calculations. The code writes
the amplitudes $A_{ph}(t)$ of Eq.~\eqref{eq11} which are calculated and stored in each time step since
for calculation of Eq.~\eqref{eq13} the entire time history is needed. This makes these calculations
extremely computation intensive.
\begin{figure}[!hpt]
	\includegraphics*[width=8cm]{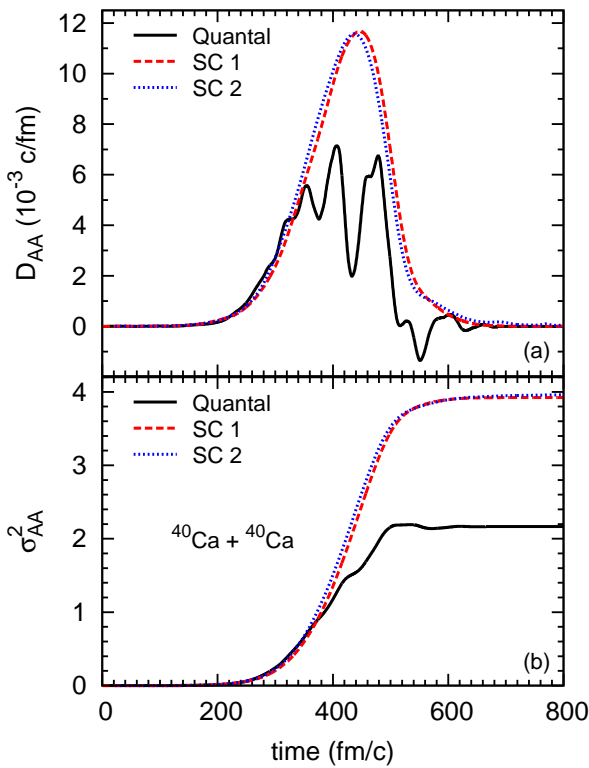}
	\caption{(Color online) Diffusion coefficient (a) and variance of fragment mass
		distribution (b) as a function of time in central collision of $^{40}$Ca + $^{40}$Ca at 52.7 MeV. Solid, dashed and dotted lines are the quantal and the semiclassical
		results with Umar~\textit{et al.}'s code and Kim~\textit{et al.}'s code, respectively.}
\end{figure} 
\begin{figure}[!hpt]
	\includegraphics*[width=8cm]{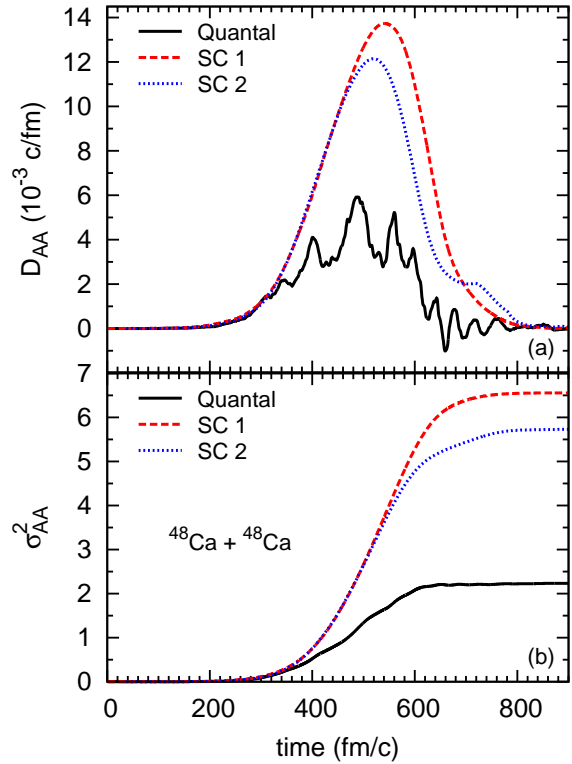}
	\caption{(Color online) Diffusion coefficient (a) and variance of fragment mass
		distribution (b) as a function of time in central collision of $^{48}$Ca + $^{48}$Ca at 50.7 MeV. Solid, dashed and dotted lines are the quantal and the semiclassical
		results with Umar~\textit{et al.}'s code and Kim~\textit{et al.}'s code, respectively.}
\end{figure} 
Formally, the unoccupied particle states consist of a finite number of negative energy bound states and an infinite number of continuum states. 
In Eq.~\eqref{eq9}, we approximate the integral over the continuum states as a sum over narrow slices (bins) in the energy space as follows,
\begin{eqnarray} \label{eq14} 
\int G_{p} (t,t' )\rho (\varepsilon _{p} )d\varepsilon _{p} \approx \sum _{j}\overline{G}_{j} (t,t') \rho _{j} \Delta \varepsilon _{j},  
\end{eqnarray} 
where the summation run over the discrete energy bins. In this expression, 
\begin{eqnarray} \label{eq15} 
\overline{G}_{j} (t,t')=\frac{1}{N_{j} } \sum _{\alpha \in \Delta \varepsilon _{j} }G_{\alpha } (t,t')  
\end{eqnarray} 
denotes the average value of the $G_{\alpha } (t,t')$ over the calculated states within the energy bin $\Delta \varepsilon _{j} $, $\rho _{j} =\rho (\varepsilon _{j} )$ is the density of states of the continuum states evaluated at the center energy $\varepsilon _{j}$ of each bin, and $N_{j}$ is the number of states in the interval.We use the Fermi gas expression for the density of states,
\begin{eqnarray} \label{eq16} 
\rho (\varepsilon _{j} )=\frac{1}{2} V\left(\frac{2m}{\hbar ^{2} } \right)^{3/2} \frac{4\pi }{(2\pi )^{3} } \sqrt{\varepsilon _{j} } =C\sqrt{\varepsilon _{j} },  
\end{eqnarray} 
where $V$ denotes the normalization volume of the continuum states. In the calculations, we use rectangular box of a volume $V=24\times 24\times 49$ fm$^{3}$, which gives  a value of $C=7.0$ MeV$^{-3/2}$ for the constant $C$. As a technical feature, in the program there is a threshold energy for the continuum positive energy proton and neutron states, $\varepsilon _{p} $ and $\varepsilon _{n} $, respectively.  Since positive energy states should begin at zero value for both protons and neutrons, in the calculations we use the level density expressions with shifted energies for protons and neutrons as follows,  
\begin{eqnarray} \label{eq17} 
\rho _{j}^{p} =\rho _{p} (\varepsilon _{j} )=C\sqrt{\varepsilon _{j} -\varepsilon _{p} },  
\end{eqnarray} 
and
\begin{eqnarray} \label{eq18} 
\rho _{j}^{n} =\rho _{n} (\varepsilon _{j} )=C\sqrt{\varepsilon _{j} -\varepsilon _{n} } . 
\end{eqnarray} 
\begin{figure}[!hpt]
\includegraphics*[width=8cm]{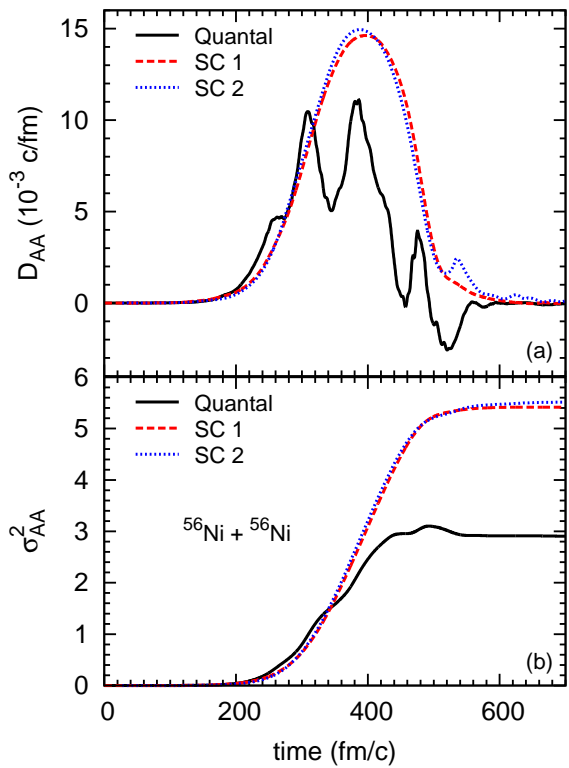}
\caption{(Color online) Diffusion coefficient (a) and variance of fragment mass
distribution (b) as a function of time in central collision of $^{56}$Ni + $^{56}$Ni at 99.9 MeV. Solid, dashed and dotted lines are the quantal and the semiclassical
results with Umar~\textit{et al.}'s code and Kim~\textit{et al.}'s code, respectively.}
\end{figure} 

We take a uniform value $\Delta \varepsilon _{j} =1.0$~MeV for the width of energy bins. The code generates finite number of discrete continuum states. Using these continuum states and hole states we calculate the diffusion nucleon coefficient $D_{AA} (t)$,
and calculate the variance of the fragment mass distribution according to,
\begin{eqnarray} \label{eq19} 
\sigma _{AA}^{2} (t)=2\int _{0}^{t}dt' D_{AA}(t') \;.
\end{eqnarray}
In principle, the variance of the fragment mass distribution should be calculated as \cite{Schroder},
\begin{eqnarray} \label{eq20} 
\sigma _{AA}^{2} (t)=\sigma _{ZZ}^{2} (t)+\sigma _{NN}^{2} (t)+2\sigma _{ZN}^{2} (t),
\end{eqnarray}
where $\sigma _{ZN}^{2}$ arises from the proton-neutron correlations in diffusion process, which is mainly driven by the symmetry energy of the binary system.
In the central collisions of symmetric systems below barrier energies, because of the relatively short collision time and small energy dissipation, 
the correlations remain small. Therefore in the calculations, we neglect the correlations and retain only the total nucleon variance given by Eq.(19).   
 
In the calculations, we gradually increase the number of discrete continuum states until the variance of fragment mass distribution reaches approximately its saturation value.  The upper panels of Fig.~(1-3)a show quantal diffusion coefficients (solid lines) for central collisions of ${}^{40}$Ca + ${}^{40}$Ca, ${}^{48}$Ca + ${}^{48}$Ca and ${}^{56}$Ni + ${}^{56}$Ni at the bombarding energies, $E_{\text{cm}} =52.7$ MeV, $E_{\text{cm}} =50.7$ MeV, and $E_{\text{cm}} =99.9$ MeV, respectively, as a function of time. The time dependence of the diffusion coefficients can also be viewed as dependence on the separation distance between ions. In the same figures, we also plot the semi-classical diffusion coefficients which are obtained with the Kim~\textit{et al.}'s code (dashed lines) and the Umar~\textit{et al.}'s code (dotted lines). The SLy4 inteaction~\cite{CB98} is employed in both codes. The reason for using both codes is to make sure that differences between the codes do not give dissimilar results. In addition to differences in numerical
procedures, Kim~\textit{et al.}'s code assumes symmetry with respect to $z=0$ plane whereas Umar~\textit{et al.}'s code does not. Furthermore Umar~\textit{et al.}'s code contains few extra time-odd terms for the Skyrme interaction~\cite{Umar05}. As we see the results from the two codes are in a reasonable agreement.
Diffusion calculations are carried out at bombarding energies slightly below the barriers. Consequently collisions do not lead to fusion in the mean-field description, after touching, the colliding ions exchange several nucleons and re-separate again. 
Overall magnitudes of the quantal diffusion coefficients are smaller than their semi-classical values and exhibits oscillations as a function of time. These oscillations in quantal calculations are partly due to the shell structure of the nuclei and partly due to the memory effect.  In fact, as a result of the non-Markovian behavior, diffusion coefficients take negative values during the separation stage of the collision. On the other hand, the semi-classical calculations exhibit a smooth behavior as a function of time or the separation distance. Part (b) in Figs. (1-3) shows the variances of the fragment mass distributions for the same systems at the same energies. The solid lines indicates the quantal results, while and the result of semi-classical calculations obtained in Kim~\textit{et al.}'s code and Umar~\textit{et al.}'s code are shown by dashed and dotted lines, respectively. The variances of the fragment mass distributions calculated in the semi-classical approximation by employing two different TDHF codes are in relatively good agreement with each other. On the other hand, the magnitude of quantal variances are smaller than the semi-classical results by nearly a factor of two in collisions of ${}^{40}$Ca + ${}^{40}$Ca and ${}^{56}$Ni + ${}^{56}$Ni, and a factor of three in ${}^{48}$Ca + ${}^{48}$Ca. This difference between the quantal and the semi-classical calculations are partly due to genuine quantal effects, shell structure and non-Markovian behavior in the diffusion coefficients. On the other hand, an important part of the difference between the quantal and the semi-classical results may be due to the density of states factor of the continuum states. In the calculations we employ the Fermi gas level density expression, which underestimates the actual density of the positive energy continuum states.

\section{Conclusions}
In this work, we investigate the nucleon exchange mechanism in the quantal framework of the SMF approach. We carry out calculations of nucleon diffusion coefficients and variances of fragment mass distributions for central collisions of ${}^{40}$Ca + ${}^{40}$Ca, ${}^{48}$Ca + ${}^{48}$Ca and ${}^{56}$Ni + ${}^{56}$Ni at the bombarding energies, $E_{\text{cm}} =52.7$ MeV, $E_{\text{cm}} =50.7$ MeV and $E_{\text{cm}} =99.9$ MeV, respectively. These bombarding energies are slightly below the fusion barriers of these systems. Consequently, colliding ions in the TDHF description do not fuse, but during contact they exchange several nucleons and separate again. In the quantal calculations we employ the TDHF code of Umar~\cite{Umar05,US91}, which is extended for obtaining time dependent particle states. We compare the quantal diffusion coefficients and the quantal variances of the fragment mass distributions with those obtained in the semi-classical framework by employing the TDHF code of Umar~\textit{et al.} and also the TDHF code of Kim~\textit{et al.} The quantal variances are smaller than those obtained in the semi-classical approximation by nearly a factor of two in collisions of ${}^{40}$Ca + ${}^{40}$Ca and ${}^{56}$Ni + ${}^{56}$Ni, and a factor of three in ${}^{48}$Ca + ${}^{48}$Ca. The difference in the results partly arises from the shell structure and non-Markovian effects in the quantal calculations. In the quantal calculations of diffusion coefficients, we use the Fermi gas expression for the level density of positive energy continuum sates. An important part in the difference between quantal and semi classical result may be due to the Fermi gas expression, which underestimates the actual level density continuum states. Further studies are needed to clarify the effect of the level density of continuum states on the quantal diffusion coefficients of nucleon exchange in heavy-ion collisions.

\begin{acknowledgments}
S.A. and S.U. gratefully acknowledge TUBITAK and Middle East technical University for partial support and warm hospitality extended to them during their visits. This work is supported in part by US DOE Grant Nos. DE-FG05-89ER40530 and DE-FG02-96ER40975, and in part by TUBITAK Grant No. 113F061.     
\end{acknowledgments}

\end{document}